\documentclass[pra,showpacs,twocolumn,floatfix]{revtex4-1}

\usepackage[utf8]{inputenc}
\usepackage{graphicx,xcolor}
\usepackage{amsmath, amsfonts, amssymb, bm}
\usepackage{slashed}

\usepackage{soul}
\def\vsigma{\vec{\sigma}}
\def\veps{\vec{\epsilon}}
\def\vex{\vec{\rm e}_x}
\def\vey{\vec{\rm e}_y}
\def\vez{\vec{\rm e}_z}

\date{\today}

\begin{document}
 \title{Controlling electron spin dynamics in bichromatic Kapitza-Dirac scattering\\ by the laser field polarization}
 \author{Matthias M. Dellweg}
 \author{Carsten M\"uller}
 \affiliation{Institut f\"ur Theoretische Physik I, Heinrich Heine Universit\"at D\"usseldorf, Universit\"atsstr. 1, 40225 D\"usseldorf, Germany}

 \begin{abstract}
 Spin-dependent Kapitza-Dirac scattering of electron beams from counterpropagating bichromatic laser waves in various polarization geometries is studied. The corresponding scattering probabilities are obtained by analytical and numerical solutions of the time-dependent Dirac equation, assuming a field frequency ratio of 2. When the fundamental field mode is circular-polarized, we show that spin dynamics are generally suppressed at low intensities, but can become distinct at high intensities. Conversely, when a linearly or elliptically polarized fundamental mode is combined with a second harmonic of circular polarization, strong spin effects arise already at low field intensities. In particular, a polarization configuration is identified which acts as a spin filter for free electrons.
 \end{abstract}

 %03.75.-b Matter waves
 %41.75.Fr Electron and positron beams (41: electron and ion optics)
 %42.50.Hz Strong-field excitation of optical transitions in quantum systems; multiphoton processes
 %42.50.Ct Quantum description of interaction of light and matter
 %42.25.Ja Polarization

 \maketitle

 \section{Introduction}
 \label{sec:introduction}
Stern and Gerlach showed in their famous experiment in 1922 that a beam of silver atoms
is transversely split into spin-up and spin-down components by an inhomogenous magnetic
field \cite{Stern-Gerlach}. After this discovery, the question was raised whether a corresponding experiment
could also be conducted with an electron beam. For charged particles, however, the
spin-separating mechanism is hindered by the influence of the Lorentz force (in combination
with the uncertainty principle), as was pointed out by Bohr and Pauli. More precisely, they
argued that it is impossible to separate an electron beam into spin components by means
of experiments based on the concept of classical particle trajectories \cite{Mott, Kessler}. 
This conclusion was later turned into a paradigm about the impossibility of a Stern-Gerlach 
spin polarizer for free electrons and has entered into the textbook literature.

First attempts to overcome this paradigm have been undertaken in the late 1990s. The
principle feasibility of separating spins in an electron beam by the inhomogenous magnetic
field of two antiparallel electric currents was analyzed in terms of classical trajectories 
\cite{Batelaan1997,Rutherford1998}. Partial splitting of spin components  by a
unidirectional inhomogenous $B$ field was shown by means of combined classical and
quantum calculations \cite{Rutherford1998a}. Complete spin separation through a
``longitudinal'' Stern-Gerlach effect, where the electron beam moves along the magnetic
field axis and is split in this direction, was obtained from a full quantum treatment
\cite{Gallup2001}.

Significant progress towards a Stern-Gerlach-like spin polarizer for electrons has been achieved
in very recent years. It has been shown theoretically that spin-polarization of electron beams
may be created by magnetic gratings which are either formed by microscopic current loops
\cite{McGregor2011} or solid-state nanostructures \cite{Tang2012}. While in the first case,
the diffraction pattern consists of characteristic spin doublet lines, in the latter case, the spin
polarization occurs a Talbot length away from the exit surface of the grating. Besides, a
Mach-Zehnder interferometer, which combines standing laser waves with static magnetic fields
from solenoids, has been proposed to exert magnetic phase control on electron beams
\cite{McGregor2011}. By exploiting fine-tuned spin-dependent phase shifts, a transverse spin
separation into two components -- similar to the original Stern-Gerlach experiment -- is predicted.
Moreover, a spin-polarizing Wien filter for electron vortex beams was described which relies on
multipolar electric and magnetic fields with cylindrical symmetry and a conversion of
orbital to spin angular momentum \cite{Karimi2012, spin-orbit}.

Very recently, two proposals have been put forward how spin polarization of free electron beams
may be attained by exploiting the Kapitza-Dirac (KD) effect \cite{Ahrens2016, Dellweg2016a}.
Previous studies of spin-dependent KD scattering have always found ``symmetric'' spin effects:
the probabilities to scatter an incident spin-up electron into a spin-down state and vice versa
coincide \cite{Ahrens2012,Batelaan_Spin,Erhard2015,Dellweg2016}. Consequently, when an unpolarized electron beam is incident, the scattered part of the beam will remain unpolarized because the spin of each single electron has flipped. This symmetry is broken in \cite{Ahrens2016} by considering an electron moving along the axis of a monochromatic standing laser wave of circular polarization. The two different spin orientations
along the beam axis are found to precess with slightly different Rabi frequencies, enabling to
divide the electron beam longitudinally into two spin-polarized portions by tailoring the interaction time.
A transverse separation of the spin components can be achieved by an interferometric arrangement of three pairs of linearly polarized laser beams \cite{Dellweg2016a}. In the first stage of this setup, the electron beam is divided into two portions by scattering from a bichromatic laser field in a three-photon KD process. Afterwards, the partial beams are superposed coherently, leading to two outgoing electron beams with a high degree of spin polarization.

The original KD effect refers to (spin-independent) electron scattering on a standing wave of light which can be formed by two counterpropagating waves of the same frequency \cite{KD}. By absorbing one photon from one of the laser beams and emitting another photon of opposite momentum into the counterpropagating beam (stimulated Compton scattering), the electron is elastically scattered. A theoretical generalization to strong laser fields has been developed \cite{Fedorov}. The effect was experimentally observed both in the Bragg and the diffraction regimes \cite{Freimund2001}. Related experiments studied KD scattering on atomic beams \cite{atoms}. For reviews on the subject, we refer to \cite{Review_Fedorov,Review_Batelaan}. Moreover, recent studies on spin-insensitive KD scattering may be found in \cite{Smirnova, Marzlin, Dellweg2015, multicolor, dielectric}.

In the present paper, we continue our investigations \cite{Dellweg2016, Dellweg2016a}
of spin-dependent KD effects in bichromatic laser fields. Our main goal is to reveal the 
influence exerted by the polarizations of the counterpropagating laser waves on the 
electron spin dynamics. To this end, the time-dependent Dirac equation is solved 
by combined analytical and numerical tools. In particular, we demonstrate pronounced
spin dynamics at relatively low laser intensities in mixed setups which combine laser 
waves of circular and linear or elliptical polarization. Instead, when both waves are 
circular-polarized, spin dynamics are shown to emerge at sufficiently high field strengths only.

Our paper is organized as follows. In Sec.~II. we perform an analytical treatment of spin-dependent bichromatic KD scattering, which is valid in the perturbative regime of low field strengths. In Sec.~III we extend our consideration to higher laser intensities by way of numerical simulations based on the time-dependent Dirac equation. Our conclusions are summarized in Sec.~IV.

For notational convenience, we set $\hbar = 1$ in this paper. The product of two four-vectors $a=(a^0,\vec{a}\,)$ and $b=(b^0,\vec{b}\,)$ is denoted as $a\cdot b=a^0b^0-\vec{a}\cdot\vec{b}$. Besides,
Feynman slash notation is employed for the four-product with Dirac $\gamma$-matrices.

 \section{Analytical considerations}
 \label{sec:analytical}
 
Throughout this paper, we shall consider the following setup. An electron of charge $-e$ is scattered from two counterpropagating laser beams, which consist of a (``strong'') right-traveling component $A_1$ with wave vector $\vec k_1 = k\vez$ and a (``weak'') left traveling component $A_2$ with wave vector $\vec k_2 = -2k\vez$. The momentum of the incident electron is chosen to lie in the $xz$-plane, i.e. $\vec{p}=(p_x,0,p_z)$, with $p_z=-2k$. By absorbing two photons from the strong mode and emitting one photon into the weak mode, the electron is Bragg scattered into the mirrored momentum state with $p_z'=2k$, thereby fulfilling the laws of energy and momentum conservation. The question is to which extent the electron spin plays a role in this scattering process, depending on the polarization geometry of the laser waves.

The physical origin of spin effects in the considered setup can be understood intuitively as follows. A three-photon process requires an interaction term which contains, in total, three powers of the external field. Hence, using the nonrelativistic terminology of Ref.~\cite{Dellweg2016}, in addition to the ${\vec A}^{\,2}$ term in the Hamiltonian (which is responsible for the original Kapitza-Dirac effect \cite{KD}), another interaction term is needed which contains the field linearly. It can be provided by the ${\vec p}\cdot{\vec A}$ term, which does not involve the spin, or the ${\vec \sigma}\cdot{\vec B}$ term. Involvement of the latter leads to spin-sensitive scattering.

The short-time behavior of the scattering probability can be obtained by a consideration within the third order of time-dependent perturbation theory. Such kind of analysis was carried out in \cite{Dellweg2016} for the case when both waves are linearly polarized along the $x$-axis. The calculation is straightforward, but rather tedious. In the same paper it was shown (see App. B therein) that the same result may be obtained from  an alternative treatment which describes the electrons by Volkov states which are dressed by the strong field mode $A_1$. The remaining effect of the weak mode $A_2$ can then be treated as a first-order perturbation. This approach has turned out to be rather convenient. It shall be followed in this section.

Regarding the KD effect as stimulated Compton scattering, we start from the $S$ matrix describing  
multiphoton Compton scattering \cite{Landau, Ehlotzky}
  \begin{equation}
   \mathcal{S} = \frac{i e}{c} \int d^4x\, \bar{\psi}_{p', s'} \slashed{A}_2 \psi_{p, s}\ .
  \label{eqn:compton_S}
  \end{equation}
Here,   
  \begin{equation}
   \psi_{p,s}(x) = \sqrt{\frac{m c}{V p^0}} \left( 1 - \frac{e \slashed{k}_1 \slashed{\mathcal{A}}_1(k_1 \cdot x)}{2 c k_1 \cdot p} \right) u_{p, s} e^{-i p \cdot x + i \Lambda_p}
  \label{eqn:volkov}
  \end{equation}
denotes the Dirac-Volkov state for the incoming electron dressed by the field $A_1$. The latter is given by a generic plane wave of the form $A_1(x) = \mathcal{A}_1(k_1 \cdot x)$ in radiation gauge, with the wave four-vector $k_1=\frac{\omega}{c}(1,\vez)$. The exponential factor contains the classical action, with the field-dependent term
  \begin{equation}
   \Lambda_p = \frac{1}{c k_1 \cdot p} \int^{k_1 \cdot x} \left[ e p \cdot \mathcal{A}_1(\phi) + \frac{e^2}{2 c} \mathcal{A}_1^2(\phi) \right] d\phi
  \label{eqn:volkov_action}
  \end{equation}
The free Dirac spinors $u_{p,s}$ are taken from \cite{Bjorken1964}, with the spin quantized along the $z$ axis. Accordingly, $\psi_{p',s'}$ represents the Dirac-Volkov state for the scattered electron.
The electron four-momenta are given by $p=(p^0, p_x, 0, -2k)$ and $p'=(p^0, p_x, 0, 2k)$, where $p^0 = \sqrt{m^2 c^2 + p_x^2 + 4 k^2}$. The momenta are assumed to be nonrelativistic so that $p^0\approx m c$.
Note that, even though we are interested in nonrelativistic electron momenta, a treatment beyond Pauli theory is necessary for laser field polarizations different from linear \cite{Erhard2015}.

We express the vector potential as $\mathcal{A}_1(\phi) = \Re \left( \epsilon_1 a_1 e^{-i \phi} \right)$,
with a polarization four-vector $\epsilon_1 = (0, \vec{\epsilon}\,)$ satisfying
$\epsilon_1^* \cdot \epsilon_1 = -1$ and $\epsilon_1 \cdot k_1 = 0$. Without loss of generality, we can choose $\epsilon_1 \cdot p$ to be real, but note that $\epsilon_1^2 = - \rho e^{i \delta}$ might still be a complex number. Its absolute value $\rho$ ranges from one for linear polarization to zero for circular polarization. A corresponding notation is employed for the counterpropagating wave, whose amplitude, polarization, wave four-vector and frequency are $a_2$, $\epsilon_2$, $k_2$ and $\omega_2=2\omega$, respectively.

The space-time integration in \eqref{eqn:compton_S} can be performed via a Fourier series expansion, according to 
   \begin{eqnarray}
    e^{i (\Lambda_p - \Lambda_{p'})}
    &=& \exp \left[ -i \left( \alpha_p -\alpha_{p'} \right) \sin(k_1 \cdot x) \right.\nonumber\\
    & & - i \rho \left( \beta_p -\beta_{p'} \right) \Re \left( i e^{i \delta} e^{- 2 i k_1 \cdot x} \right) \nonumber\\
    & & \left. - 2 i \left( \beta_p - \beta_{p'} \right) k_1 \cdot x \right] \nonumber\\
    &=& \sum_{n \in \mathbb{Z}} \tilde{J}_n \left( \alpha_p - \alpha_{p'}, \rho \left( \beta_p - \beta_{p'} \right); e^{i\delta} \right) \nonumber\\
    & & \times e^{-i n k_1 \cdot x - 2 i \left( \beta_p - \beta_{p'} \right) k_1 \cdot x}\ ,
  \label{eqn:S_phase}
  \end{eqnarray}
with the abbreviations
  \begin{equation}
    \alpha_p = -\frac{e a_1}{c} \frac{p \cdot \epsilon_1}{k_1 \cdot p}\ ,
    \qquad
    \beta_p = \frac{e^2 a_1^2}{8 c^2} \frac{1}{k_1 \cdot p}\ .
  \label{eqn:volkov_abbr}
  \end{equation}
Besides, the generalized Bessel functions
  \begin{equation}
  \tilde{J}_n(u, v; t) = \sum_{\ell \in \mathbb{Z}} J_{n-2\ell}(u) t^\ell J_\ell(v) \ ,
  \end{equation}
with the ordinary Bessel functions $J_\ell$, have been used.  
  
In correspondence, the $S$-matrix for the electron transition from $p$ to $p'$ by absorbing two photons from $A_1$ and emitting one into $A_2$ can be expressed as 
  \begin{eqnarray}
    \mathcal{S} &\approx& \frac{ie}{cV} \int d^4x\, \bar{u}_{p', s'} \Bigg(
     \slashed{A}_2^{(+)} \tilde{J}_2 e^{i \left( p' - p - 2k_1 \right) \cdot x} \nonumber\\
     & & \left. - \frac{e}{2 c} \left[ \frac{\slashed{A}_1^{(-)} \slashed{k} \slashed{A}_2^{(+)}}{k_1 \cdot p'} + \frac{\slashed{A}_2^{(+)} \slashed{k}_1 \slashed{A}_1^{(-)}}{k_1 \cdot p} \right]
     \tilde{J}_1 e^{i \left( p' - p - k_1 \right) \cdot x}
     \right) u_{p, s} \nonumber\\
     &\approx& \frac{i e}{2} T \bar{u}_{p', s'} \left[
      a_2 \tilde{J}_2 \bar{\slashed{\epsilon}}_2
     - \frac{e a_1 a_2}{4 c} \tilde{J}_1 \left( \frac{\slashed{\epsilon}_1 \slashed{k}_1 \bar{\slashed{\epsilon}}_2}{k_1 \cdot p'}
      + \frac{\bar{\slashed{\epsilon}}_2 \slashed{k}_1 \slashed{\epsilon}_1}{k_1 \cdot p} \right)
     \right] u_{p, s} \nonumber\\
  \label{eqn:S1}
  \end{eqnarray}
Here, $\slashed{A}_1^{(-)} = \frac{1}{2} a_1 \slashed{\epsilon}_1 e^{-i k_1 \cdot x}$ defines the component of  $\slashed{A}_1$ which describes absorption of one photon from this wave. Similarly, $\slashed{A}_2^{(+)} = \frac{1}{2} a_2 \bar{\slashed{\epsilon}}_2 e^{i k_2 \cdot x}$ describes one-photon emission into $A_2$, where $\bar{\slashed{\epsilon}}_2=\epsilon_2^*\cdot\gamma$. For notational simplicity, the arguments of the generalized Bessel functions have been omitted. Note that in Eq.~\eqref{eqn:S1} we have restricted ourselves to the resonant contribution, which agrees with the Bragg condition. The integration has thus produced a factor of $c V T$, with the formal interaction time $T$ and quantization volume $V$.

The spinor-matrix products in Eq.~\eqref{eqn:S1} read
  \begin{equation}
    \left( \bar{u}_{p', s'} \bar{\slashed{\epsilon}}_2 u_{p, s} \right)_{s', s}
    = -\frac{p_x}{m c} \vec{\epsilon}_2^{~*} \cdot \vec{e}_x + i \frac{2 \omega}{m c^2} \left( \vec{\epsilon}_2^{~*} \times \vec{e}_z \right) \cdot \vsigma  
  \label{eqn:matrix1}
  \end{equation}  
and 
  \begin{eqnarray}
    & & \left( \bar{u}_{p',s'}\left[
     \frac{\slashed{\epsilon}_1 \slashed{k}_1 \bar{\slashed{\epsilon}}_2}{k_1 \cdot p'} + \frac{\bar{\slashed{\epsilon}}_2 \slashed{k}_1 \slashed{\epsilon}_1}{k_1 \cdot p}
    \right] u_{p,s} \right)_{s', s} \nonumber\\
    &\approx& \frac{2 \vec{\epsilon}_1 \cdot \vec{\epsilon}_2^{~*}}{m c} 
    + \frac{4 i \omega}{m^2 c^3} \left( \vec{\epsilon}_1 \times \vec{\epsilon}_2^{~*} \right)\cdot \vsigma
  \label{eqn:matrix2}
  \end{eqnarray}
The notation $(\ldots)_{s,s'}$ on the left-hand sides in Eqs.~\eqref{eqn:matrix1} and \eqref{eqn:matrix2} is meant to indicate $2\times2$ matrices whose entries are indexed by the spin quantum numbers.
  
Furthermore, from the Taylor series of the generalized Bessel functions, we obtain
  \begin{eqnarray}
   \tilde{J}_1
    &\approx& \frac{\alpha_p - \alpha_{p'}}{2}
    \approx -2 \frac{e a_1}{m c^2} \frac{p_x}{m c} \veps_1 \cdot \vex \nonumber\\
   \tilde{J}_2
     &\approx& \frac{\left( \alpha_p - \alpha_{p'} \right)^2}{8}
     + \vec{\epsilon}_1^{~2} \frac{\beta_p - \beta_{p'}}{2} \nonumber\\
     &\approx& \frac{e^2 a_1^2}{m^2 c^4} \left( 2 \frac{p_x^2}{m^2 c^2} \left( \vec{\epsilon}_1 \cdot \vex \right)^2 - \frac{1}{4} \veps_1^{~2} \right)
  \end{eqnarray}
The applicability of these approximations requires, in particular, that the momentum component $p_x$ is sufficiently small.
  
Putting all pieces together, the $S$-matrix adopts the form
  \begin{eqnarray}
    \mathcal{S} \approx \frac{i}{2} T \frac{e^3 a_1^2 a_2}{m^3 c^6} \hat{\Omega}
    = \frac{i}{2} T \xi_1^2 \xi_2 \hat{\Omega}\ ,
  \label{eqn:S2}
  \end{eqnarray}
with the spin- and polarization-dependent Rabi matrix 
  \begin{eqnarray}
  \hat{\Omega} &:=& \frac{1}{4} p_x c \left( \veps_1^{~2}\, \vec{\epsilon}_2^{~*} \cdot \vex
      + 4 \, \veps_1 \cdot \vex\, \veps_1 \cdot \vec{\epsilon}_2^{~*} \right)  \nonumber\\
      & & - \frac{i}{2} \omega \veps_1^{~2} \left( \vec{\epsilon}_2^{~*} \times \vez \right) \cdot \vsigma\ .
  \label{eqn:matrixR}    
  \end{eqnarray}
Besides, in the second step, the dimensionless field amplitudes $\xi_{1, 2} := \frac{e a_{1, 2}}{m c^2}$, which are commonly used in strong-field atomic physics, have been introduced. Equation \eqref{eqn:S2} represents the perturbative limit of the three-photon KD process, which assumes $\xi_1, \xi_2 \ll 1$.

The polarization states of the counterpropagating laser waves enter through the Rabi matrix. For various combinations of linearly, circularly and elliptically polarized waves, the specific form of $\hat\Omega$ is indicated in Table \ref{tab:S_matrix}. 

  \begin{table}
   \begin{tabular}{rcl|cc|c}
    $A_1$ & -- & $A_2$ & $\vec{\epsilon}_1$ & $\vec{\epsilon}_2$ & $\hat{\Omega}$ \\
    \hline
    lin. & -- & lin. & $\vex$ & $\vex$ & $\frac{5}{4} p_x c + \frac{i}{2} \omega \sigma_y$ \\
    circ. & -- & circ. & $\frac{1}{\sqrt{2}}( \vex + i \vey )$ & $\frac{1}{\sqrt{2}}( \vex + i \vey )$ & $\frac{1}{\sqrt{2}}p_x c$ \\
    circ. & -- & c-circ. & $\frac{1}{\sqrt{2}}( \vex + i \vey )$ & $\frac{1}{\sqrt{2}}( \vex - i \vey )$ & $0$ \\
    circ. & -- & lin. & $\frac{1}{\sqrt{2}}( \vex + i \vey )$ & $\vex$ & $\frac{1}{2}p_x c$ \\
    lin. & -- & circ. & $\vex$ & $\frac{1}{\sqrt{2}}( \vex + i \vey )$ & $\frac{5}{4 \sqrt{2}}p_x c - \frac{\omega}{2 \sqrt{2}} \sigma_-$ \\
    ell. & -- & circ. & $\frac{1}{\sqrt{26}}( \vex + 5 i \vey )$ & $\frac{1}{\sqrt{2}}( \vex + i \vey )$ & $\frac{6}{13 \sqrt{2}} \omega \sigma_-$ \\
    \hline
    \hline
   \end{tabular}
   \caption{\label{tab:S_matrix}
    Specific forms of the Rabi matrix $\hat\Omega$ [see Eq.~\eqref{eqn:matrixR}] for various polarizations configurations of the laser waves. The abbreviation "c-circ." stands for counterrotating circular polarization. The usual definition $\sigma_\pm := \sigma_x \pm i \sigma_y$ is applied.
   }
  \end{table}

 \section{Numerical Results and Discussion}
 \label{sec:results}

In order to confirm our analytical predictions and to extend the consideration to higher field intensities  beyond the leading order of perturbation theory, we have conducted numerical simulations. 
The time-dependent Dirac equation
  \begin{equation}
   \left( i \slashed{\partial} + \frac{e}{c} \slashed{A}(x) - mc \right) \psi(x) = 0
  \label{eqn:dirac}
  \end{equation} 
was solved in the presence of the combined laser fields $A(x)=f(t)[\mathcal{A}_1(k_1\cdot x)+\mathcal{A}_2(k_2\cdot x)]$. A slowly varying envelope function $f(t)$ is introduced here to model the switching on and off of the laser field in the simulation. The turn-on and turn-off phases are sin$^2$-shaped and comprise five cycles of the fundamental laser period.
Due to the periodicity of the external field, the electron wave function can be written as a discrete expansion into momentum eigenstates of the form
  \begin{equation}
   \psi(x) = \sum_{n\in\mathbb{Z}}\sum_{s\in\{\uparrow,\downarrow\}}
   \left[ c_n^s(t) u_{p_n,s} + d_n^s(t) v_{p_n,s} \right]\,e^{i\vec{p}_n\cdot \vec{x}}\ ,
  \label{eqn:ansatz}
  \end{equation} 
with the free Dirac spinors $u_{p,s}$ and $v_{p,s}$ from \cite{Bjorken1964} and $\vec{p}_n=(p_x,0,nk)$. Inserting this ansatz into the Dirac equation \eqref{eqn:dirac} yields a coupled system of ordinary differential equations for the time-dependent expansion coefficients $c_n^s(t)$ and $d_n^s(t)$, with $n\in\mathbb{Z}$ and $s\in\{\uparrow,\downarrow\}$ \cite{projection}. We note that the same kind of approach was used in \cite{Dellweg2016} within Pauli theory for bichromatic laser fields and in \cite{Ahrens2012} within Dirac theory for a standing laser wave. The ``positronic'' contributions in Eq.~\eqref{eqn:ansatz} are required to obtain a complete set of basis states with momentum $\vec{p}_n$ for the numerical propagation of the wave function in the presence of the laser fields. The initial and final states, though, when the laser field is switched off, will contain only electronic contributions with corresponding coefficients $c_n^s(t)$. While the coupled system of equations formally has an infinite dimension, in our simulations it has been truncated to a finite number of momentum modes, $-n_{\rm max}\le n\le n_{\rm max}$, with $n_{\rm max}$ being large enough to reach convergence. The finite system of equations was solved by Runge-Kutta methods.

This way, the time evolution of the spinor wave function of the electron is obtained, starting from an initial state with longitudinal momentum $p_z=-2k$ and spin projection either up ($s=\ \uparrow$) or down ($s=\ \downarrow$). Below, these initial states and the corresponding scattered states with $p_z'=2k$ shall be symbolically denoted as $|-2,\uparrow\rangle$, $|-2,\downarrow\rangle$, $|2,\uparrow\rangle$, and $|2,\downarrow\rangle$. 
Their probability amplitudes are given by $c_{-2}^\uparrow(t)$, $c_{-2}^\downarrow(t)$, $c_{2}^\uparrow(t)$, and $c_{2}^\downarrow(t)$, respectively, in accordance with Eq.~\eqref{eqn:ansatz}.
In the regime of laser parameters under consideration, the outgoing electron state after the interaction with the laser fields can be expanded into these four states. While states with other longitudinal momenta are important during the interaction and taken into account in the simulation, their contributions to the final electron state are neglibibly small. Note besides that the value of the initial transverse momentum $p_x$ is suppressed in the symbolic notation. This is justified because the canonical momentum in this direction is conserved, so that $p_x' = p_x$.

For reasons of practical feasibility, our numerical computations assume high-intensity laser fields with frequencies in the x-ray domain. Radiation sources with corresponding characteristics are, in principle, available through high-harmonic emission from plasma surfaces \cite{plasmaHHG} or high-power x-ray free-electron laser (XFEL) facilities, such as the European XFEL (Hamburg, Germany) or the Linac Coherent Light Source (Stanford, California) \cite{XFEL}. At present, the latter is able to generate brilliant x-ray pulses with up to $\sim 10$\,keV photon energy, $\sim 10^{20}$\,W/cm$^2$ intensity, and $\sim 100$\,fs duration.

\subsection{Both modes of linear polarization}

The scenario, when both waves are linear-polarized along the $x$ axis, was considered within Pauli theory in \cite{Dellweg2016}, focussing on incident electron momenta with $p_x=0$. In this case, the Rabi frequency $\Omega_R = \frac{1}{2}\xi_1^2\xi_2\omega$ results from Eq.~\eqref{eqn:S2}, in agreement with Eq.~(16) in \cite{Dellweg2016}. This frequency describes the flopping dynamics between the resonantly coupled momentum states $p$ and $p'$.

\begin{figure}[b]
\begin{center}
\includegraphics[width=0.45\textwidth]{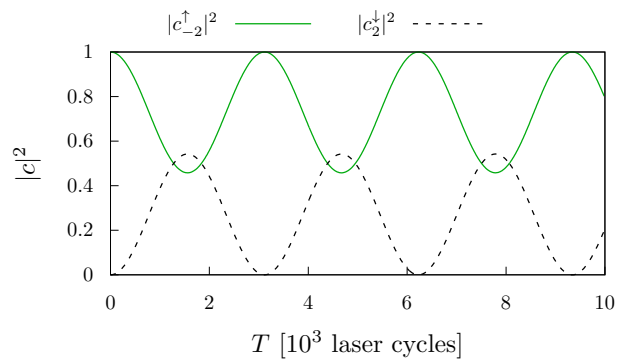}
\end{center}
\caption{Final occupation probabilities $|c_{-2}^\uparrow(T)|^2$ (green solid line) 
and $|c_{2}^\downarrow(T)|^2$ (black dashed line) for KD scattering from 
bichromatic counterpropagating laser waves of linear polarization, 
as a function of the interaction time $T$. The field parameters are 
$\omega = 2\times 10^3$\,eV, $ e a_1 = 5.656 \times 10^4$\,eV and 
$e a_2 = 2 \times 10^4$\,eV, corresponding to a combined laser intensity 
of $I=6.54\times 10^{22}\,\mathrm{W/cm^2}$. The electron is incident with 
longitudinal momentum of $p_z= -2k$ and transverse momentum of $p_x = 0$.}
\label{lin-lin}
\end{figure}

As an illustration, the outcome of a numerical simulation for this polarization geometry is depicted in Fig.~\ref{lin-lin}. It  shows the dependence of the final-state electron population on the interaction time, i.e., every plotted data point corresponds to a full interaction with switching on and off the fields.
A Rabi oscillation between the states $|-2,\uparrow\rangle$ and $|2,\downarrow\rangle$ takes place. The amplitude $C$ of these oscillations does not reach the maximal value of 1, though. The population of the scattered state rather follows the behavior
\begin{equation}
|c_{2}^\downarrow(T)|^2 = C\,\sin^2\left( \frac{1}{2}\Omega T \right)\ ,
\label{C}
\end{equation}
with the oscillation frequency $\Omega = \Omega_R / \sqrt{C}$. The fact that the Rabi oscillations are not fully developed can be attributed to an intrinsic detuning which arises when the field intensities are not so small.
This phenomenon was discussed in Ref.~\cite{Dellweg2016} and related to the asymmetric field configuration, due to which the four-momenta of the incoming and the outgoing electron, respectively, may be subject to uneven dressing effects (see Sec.~III.B and App.~A therein). Hence, perceptible mismatches in the energy-momentum balance of the scattering process, which occurs inside the field, can arise.

The oscillation amplitude in Fig.~\ref{lin-lin}, following from Dirac theory, amounts to $C\approx 0.54$. We point out that a slightly smaller amplitude of $C\approx 0.51$ would result from a numerical solution of Pauli's equation. This indicates that some relativistic effects are present in the detuning which arises for the parameters in Fig.~\ref{lin-lin}.

We emphasize that the same kind of Rabi oscillation would arise if the electron was incident with opposite spin instead. Then the flopping dynamics would occur between the states $|-2,\downarrow\rangle$ and $|2,\uparrow\rangle$. The corresponding scattering probabilities for up$\to$down and down$\to$up are the same in the scenario where both laser waves are linear-polarized, as was mentioned in the Introduction. The other polarization geometries shall be discussed in the subsequent sections.

\subsection{Strong mode of circular polarization}

When studying spin effects in KD scattering, it appears natural to consider fields of circular polarization. In this case, the photons carry a definite spin along the beam axis (helicity) which might be transfered to the electron during the combined emission-absorption process. Indeed, it has recently been shown that pronounced spin dynamics may arise in two-photon KD scattering processes which occur in a monochromatic standing laser wave of circular polarization \cite{Erhard2015,Ahrens2016}; see also \cite{Bauke_Spin} for related studies. In general, however, spin-flip transitions in circular-polarized fields have to compete with spin-preserving transitions because -- in the nonrelativistic language of Pauli theory -- the $\vec{p} \cdot \vec{A}$ interaction term cannot be avoided. As a result, spin flips are expected to be suppressed, unless the electron moves along the laser beam axis \cite{Erhard2015,Ahrens2016}.

In this subsection, we analyze spin effects in bichromatic KD scattering when the strong mode $A_1$ is circularly polarized. It has been speculated that distinct spin effects could arise when an electron is KD scattered from bichromatic laser fields which both are circularly polarized \cite{Freimund2003}. The underlying idea is that, in a three-photon process, the net absorption of spin angular momentum from the field can be one unit of $\hbar$. This spin transfer could be compensated by a spin flip of the electron along the laser beam axis. While this idea is very appealing, the classical simulations, which were carried out in \cite{Freimund2003} based on the Bargman-Michell-Telegdi equation, could not reveal spin flips, though.

Our analytical calculations confirm that, to leading order, there is no spin flip involved in the three-photon KD effect. Rather, as shown in Table \ref{tab:S_matrix}, the Rabi matrix is given by $\hat\Omega = \frac{1}{\sqrt{2}}p_x c$ and thus independent of the spin. This results holds for two counterpropagating waves which are circularly polarized in the same sense. According to Table \ref{tab:S_matrix}, qualitatively the same  conclusion holds, when only the strong wave is circular-polarized, whereas the weak mode is linear-polarized. When, instead, counterrotating circular-polarized laser waves are applied, no scattering occurs at all. This is in line with the outcome of \cite{Erhard2015} regarding two-photon KD processes.

\begin{figure}[h]
\begin{center}
\includegraphics[width=0.45\textwidth]{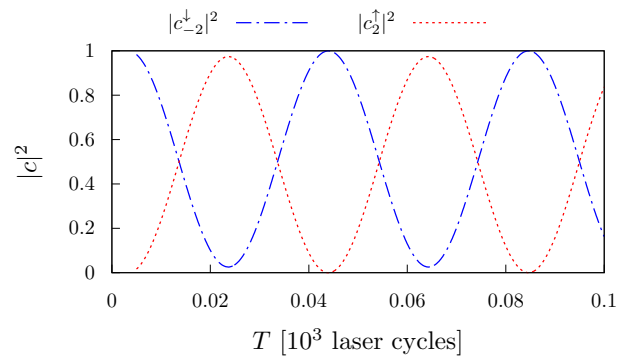}
\end{center}
\caption{Final occupation probabilities $|c_{-2}^\downarrow(T)|^2$ (blue dash-dotted line) 
and $|c_{2}^\uparrow(T)|^2$ (red dotted line) for KD scattering from 
bichromatic counterpropagating laser waves with corotating circular polarization, 
as a function of the interaction time $T$. 
The field parameters are $\omega = 2\times 10^3$\,eV,
$e a_1 = e a_2 = 8 \times 10^4$\,eV. The combined laser intensity is 
$I=4.36\times 10^{23}\,\mathrm{W/cm^2}$. The initial electron momentum
components are the same as in Fig.~~\ref{lin-lin}.}
\label{circ-circ}
\end{figure}

Nevertheless, the analytical predictions of the previous section only apply to the low-intensity regime. By performing a systematic series of numerical simulations, we found that pronounced spin dynamics in corotating circular-polarized fields may arise in a specific range of parameters. An example is shown in Fig.~\ref{circ-circ}. For an incident electron state $|-2,\downarrow\rangle$ with zero transverse momentum, a Rabi oscillation dynamics with the $|2,\uparrow\rangle$ state appears. Remarkably, the electrons are only scattered when they are initially in the spin-down state, but not when they are in the spin-up state. The effective scattering term must therefore be proportional to $\sigma_+$. We note besides that the measured oscillation frequency does not follow the $\xi_1^2\xi_2$--scaling of Eq.~\eqref{eqn:S2}. Our results therefore suggest that the scattering occurs in a nonperturbative domain, is driven highly non-linear, and the number of involved photons cannot be discriminated easily as before due to substantial detuning.
  
\begin{figure}[h]
\begin{center}
\includegraphics[width=0.4\textwidth]{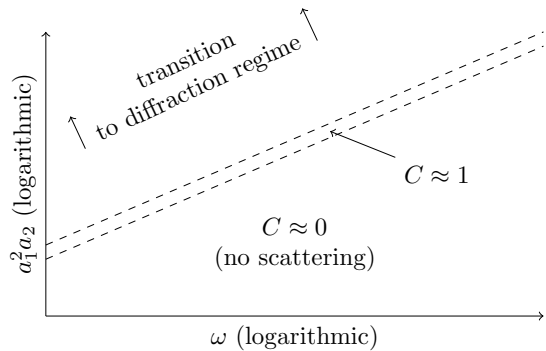}
\end{center}
\caption{Schematic diagram of the maximal Rabi amplitude $C$ [see Eq.~\eqref{C}]
in the $a_1^2a_2$--$\omega$ plane, indicating various parameter regimes, 
when both waves are corotating circular-polarized. The dyke summit 
is located approximately at $e^3 a_1^2a_2 \sim \omega^2 10^6$\,eV
(order of magnitude).}
\label{dyke-circ}
\end{figure}

A summary of our parameter scan is illustrated by the schematic diagram in Fig.~\ref{dyke-circ}. In analogy with the case when both fields are linear-polarized (see Fig.~7 in \cite{Dellweg2016}), a dykelike structure separates various interaction regions. In the current case of two corotating circular-polarized fields, however, in the perturbative region far below the dyke no spin-dependent KD scattering is found. Rabi oscillations with sizeable amplitude rather occur in the region just above the dyke, where the field intensities are relatively high.

We point out that spin-dependent KD scattering in circular-polarized (or, more generally, elliptical-polarized) laser fields requires a theoretical treatment beyond the Pauli equation, as was shown in Ref.~\cite{Erhard2015}. Being based on the Dirac equation, our present considerations fully account for relativistic effects. The leading-order spin-dependent relativistic corrections to the Pauli Hamiltonian are of the form $\vec{\sigma}\cdot\left(-\frac{i\hbar}{mc}{\vec E}\times\vec{\nabla}\right)$ and ${\vec \sigma}\cdot\left(\frac{e}{mc^2}{\vec E} \times {\vec A}\right)$, representing spin-orbit interaction and the coupling of the electron spin to the photonic spin density of the laser waves, respectively \cite{Erhard2015}. Here, ${\vec E}=-\frac{1}{c}\dot{\vec A}$ denotes the laser electric field. For the parameters of our numerical simulations in Fig.~\ref{circ-circ}, the spin-orbit term is rather small. In comparison with the ${\vec \sigma}\cdot{\vec B}$ term, it is suppressed by a factor $\frac{p}{mc}\sim 10^{-2}$. Relatively more important is the spin density term since $\frac{ea_{1,2}}{mc^2}\sim 10^{-1}$ in our numerical example.

\subsection{Weak mode of circular polarization}

In the previous subsection we have seen that a strong (i.e. fundamental) field mode 
of circular polarization is only beneficial for spin effects in bichromatic KD scattering
when the field intensities are sufficiently high. 
When, instead, the weak (i.e. second-harmonic) field mode is circularly polarized, 
the analytical considerations in Sec.~II. predict pronounced spin dynamics already 
at low laser intensities where perturbation theory applies.

\begin{figure}[h]
\begin{center}
\includegraphics[width=0.45\textwidth]{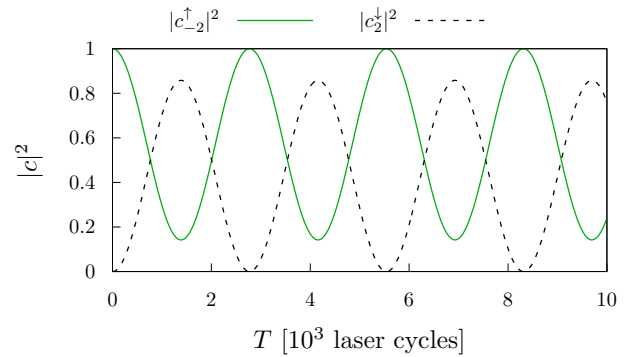}
\end{center}
\caption{Final occupation probabilities $|c_{-2}^\uparrow(T)|^2$ (green solid line) 
and $|c_{2}^\downarrow(T)|^2$ (black dashed line) for KD scattering from a laser field
consisting of a linear-polarized wave with $\omega_1 = 2\times 10^3$\,eV, 
$ e a_1 = 5.656 \times 10^4$\,eV and a counterpropagating circular-polarized
wave with $\omega_2 = 4\times 10^3$\,eV, $e a_2 = 2 \times 10^4$\,eV.
The combined laser intensity is $I=6.54\times 10^{22}\,\mathrm{W/cm^2}$.
The electron is incident with longitudinal momentum of $p_z= -2k$ and 
transverse momentum of $p_x = 0$.}
\label{lin-circ-0}
\end{figure}

This prediction is confirmed by a numerical solution of the Dirac equation. 
Figure~\ref{lin-circ-0} shows the Rabi oscillations between an incident electron
state $|-2,\uparrow\rangle$ and the scattered spin-flipped state $|2,\downarrow\rangle$
when the weak mode is circular-polarized and the strong-mode linear-polarized. 
A vanishing transverse electron momentum is assumed. Because of detuning effects 
\cite{Dellweg2016}, the oscillations are not fully developed but reach a large 
amplitude of $C\approx 0.8$. 

Note that, when the electron was initially in state $|-2,\downarrow\rangle$, no
spin-flipping transitions are found in our simulations. In fact, the electron will
remain in its initial state, then. This is in accordance with the Rabi matrix
being proportional to $\sigma_-$, see Table~I. Thus, in contrast to the case of
two linearly polarized laser waves in Sec.~III.A, the symmetry between the up$\to$down and 
down$\to$up transition probabilities is broken in the present field configuration \cite{sigma-plus}.

\begin{figure}[h]
\begin{center}
\includegraphics[width=0.45\textwidth]{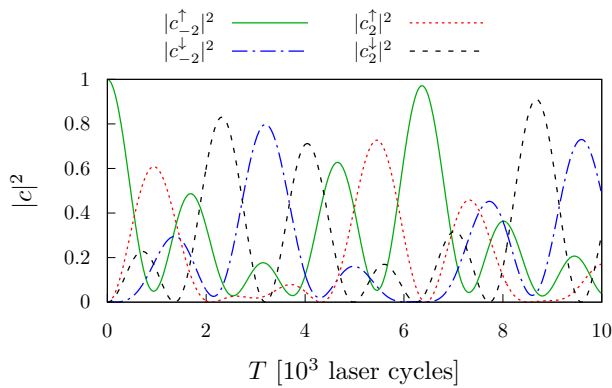}
\end{center}
\caption{Same as Fig.~\ref{lin-circ-0} but with $p_x = k$. 
Due to the nonzero transverse momentum, also the electron states
$|2,\uparrow\rangle$ and $|-2,\downarrow\rangle$ are populated. 
The corresponding occupation probabilities are shown by the 
red dotted and blue dash-dotted lines, respectively.}
\label{lin-circ-px}
\end{figure}

The clean picture of a two-state flopping dynamics changes dramatically
when the electron is incident with nonzero transverse momentum. An example
is depicted in Fig.~\ref{lin-circ-px}. Now the electron probability density 
is distributed over all four states, leading to a rather complex evolution 
with the interaction time. One can see that, in the limit of small $T$, 
the initial population is transfered not only into the spin-flipped 
$|2,\downarrow\rangle$ state but also into the spin-preserving 
$|2,\uparrow\rangle$ state. This is because the Rabi matrix contains
an additional term now, which is spin-independent and proportional to $p_x$
(see Table~I). Due to this term, population can also be shifted from the 
$|2,\downarrow\rangle$ state to the $|-2,\downarrow\rangle$ state, which
accordingly starts to occur with some time delay, as Fig.~\ref{lin-circ-px}
illustrates. 

In a real experiment, the detrimental effect caused by nonzero $p_x$ can
hardly be avoided because the incident electron beam will always have some
transverse momentum width. In light of this, the combination of linear and
circular polarized laser waves is not yet optimal.

Interestingly, the robustness of the setup can be largely enhanced,
when a small degree of ellipticity is imprinted on the strong mode.
By a suitable choice of the polarization vectors, the $p_x$-dependent
term in the Rabi matrix can be removed [see Eq.~\eqref{eqn:matrixR}].
This is achieved, for example, when the strong mode possesses an
ellipticity degree of 20\%. As indicated in Table~I, for this 
specially tailored field geometry, the Rabi matrix becomes 
proportional to $\sigma_-$, even for nonvanishing values of $p_x$.  

\begin{figure}[b]
\begin{center}
\includegraphics[width=0.45\textwidth]{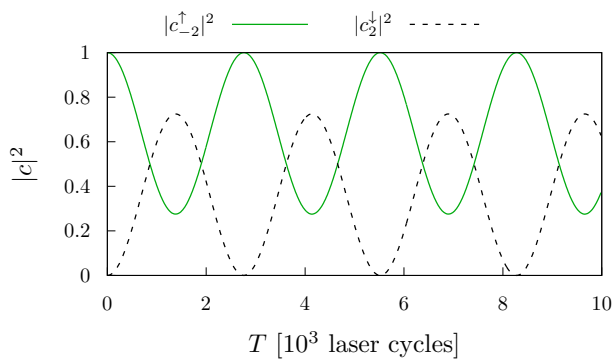}
\end{center}
\caption{Same as Fig.~\ref{lin-circ-px}, but the fundamental laser
wave is elliptically polarized with polarization vector given by
$\vec{\epsilon}_1=\frac{1}{\sqrt{26}}( \vex + 5 i \vey )$.
As a consequence, only the states $|-2,\uparrow\rangle$ and 
$|2,\downarrow\rangle$ couple with each other.}
\label{ell-circ}
\end{figure}

An example of the population dynamics in the optimized setup is shown in Fig.~\ref{ell-circ}, 
with $p_x=k$. While the same value of transverse momentum led to a very complicated
behavior in Fig.~\ref{lin-circ-px}, in the present case the detrimental impact of $p_x$ 
is balanced by the properly chosen ellipticity of the strong mode. 
As a result, a clean Rabi oscillation between the two states 
$|-2,\uparrow\rangle$ and $|2,\downarrow\rangle$ is obtained, 
reaching an amplitude of about $C\approx 0.7$. We verified numerically 
that the interaction in this field configuration is indeed independent of $p_x$, 
at least for $p_x\lesssim 30k$ and the other parameters chosen as in Fig.~\ref{ell-circ}.

Note that, when the electron was initially in state $|-2,\downarrow\rangle$, no
spin-flipping transitions are found in our simulations.

In the context of the current section, a recent study on pair production 
in very strong laser fields is worth mentioning. As was shown in \cite{Wollert}, 
spin-polarized electron-positron pairs can be created in two 
counterpropagating, monochromatic laser beams of elliptical polarization.

\subsection{Spin polarization in linear-polarized fields}
Spin-dependent three-photon KD scattering in bichromatic fields, which both are linearly polarized, was discussed in \cite{Dellweg2016}. As already mentioned above, symmetric spin effects arise in this case, where the scattering probabilities for up$\to$down and down$\to$up are the same. 

\begin{figure}[t]
 \begin{center}
\includegraphics[width=0.4\textwidth]{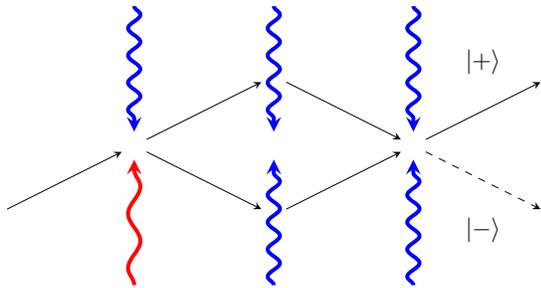}
 \caption{\label{fig:skizze}
  Scheme of the spin-polarizing interferometric beam splitter, as introduced in \cite{Dellweg2016a}.
  An incident electron beam undergoes three stages of KD scattering. As a result, 
  the outgoing electron beam is separated into its spin components $|+\rangle$ and $|-\rangle$ along the laser magnetic field, in close analogy to the Stern-Gerlach effect.}
\end{center}
\end{figure}

Nevertheless, a spin polarizer can be formed when three pairs of linearly polarized laser fields are combined in an interferometric setup. Then, one is sensitive to the quantum phase generated by the spin-dependent bichromatic KD effect, which forms the first step in the spin-polarizing beam splitter (see Fig.~\ref{fig:skizze}). The setup may be viewed as three consecutive ponderomotive gratings, from which the incident electron beam diffracts. 
They are given by the formulas
  \begin{equation}
   V_\textrm{b} (z) = - \frac{e^3 a_1^2 a_2 \omega}{2 m^3 c^6} \sin(4 k z) \sigma_y
  \label{eqn:bi_pot}
  \end{equation}
for a bichromatic wave (first stage of the interferometer) and
\begin{equation}
   V_\textrm{m} (z) = \frac{e^2 a_0^2}{8 m c^2} \cos \left( 4 k z + \chi \right)
  \label{eqn:mono_pot}
  \end{equation}
for a monochromatic standing laser wave (second and third stage of the interferometer). 
Upon scattering from the spin-dependent potential \eqref{eqn:bi_pot}, the incident electron beam is split
into two portions, whose spin and momentum are entangled. Afterwards, both partial beams are reflected by a first interaction with the spin-independent potential \eqref{eqn:mono_pot} and, finally, coherently mixed by another interaction with a monochromatic standing laser wave. As a result, the outgoing electron beam is split into its spin components along the laser magnetic field.

Details on the derivation of the spin-dependent effective ponderomotive potential in Eq.~\eqref{eqn:bi_pot} are provided in \cite{Dellweg2016, Dellweg2016a}. It relies on a Magnus expansion of the corresponding Pauli-Hamiltonian. We would like to point out that the potential given in Eq.~(20) of \cite{Dellweg2016} contains an errorenous sign. In this form, the equation would hold for a negative value of the unit charge $e$. The mistaken sign, though, does not affect the results presented there.

\begin{figure}[b]
\begin{center}
\includegraphics[width=0.48\textwidth]{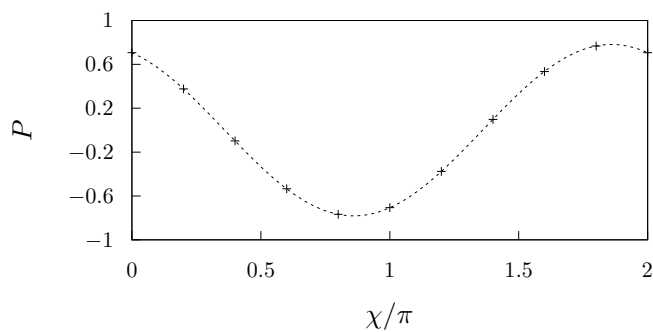}
\end{center}
\caption{Dependence on the phase parameter $\chi$ [see Eq.~\eqref{eqn:mono_pot}] of the polarization degree of the outgoing electron beam with positive momentum, as obtained by the spin-polarizing interferometric beam splitter of Ref.~\cite{Dellweg2016a}. The black crosses present the numerical data, whereas the dotted line is an analytical curve of the form $P_0\sin(\chi + \delta)$, fitted over the parameters $P_0$ and $\delta$. The outgoing portion of the electron beam with negative momentum has the opposite degree of polarization.}
\label{degree}
\end{figure}

The relative position of the standing wave nodes can be adjusted by the phase shift parameter $\chi$ in the  potential \eqref{eqn:mono_pot}. It turns out that the choice of this parameter has a strong impact on the performance of the spin polarizer. In \cite{Dellweg2016a}, the phase parameter was chosen as $\chi = -\frac{\pi}{10}$ in order to reach an optimized polarization degree of 77\% for each of the outgoing beam portions. In Fig.~\ref{degree} we show how the polarization degree depends on this parameter.

Our findings, that the polarization degree $P$ strongly depends on $\chi$ and that the maximally 
achievable value of $P$ remains below 100\%, may be understood qualitatively as follows. 
The field-induced detuning arising in the bichromatic KD effect has a twofold negative impact 
on the performance of the beam splitter. First, the reduced Rabi oscillation amplitude
leads to an uneven splitting of the beam in the first stage. This results in imperfect
polarization because the destructive interference in the final stage for either spin states
cannot be complete. 
Second, because of the intrinsic detuning, the (mean) energies of the incident momentum mode 
and the scattered momentum mode slightly differ during the interaction. As a consequence, 
a relative offset developes between the quantum phases accumulated by both modes.
This is harmful for the spin polarization achieved after their coherent 
superposition in the final stage of the interferometer. We argue that, by a proper choice 
of $\chi$, this second detrimental effect can be compensated because $\chi$ enters into 
the same quantum phases. This is why the resulting degree of polarization is modulated 
by the phase parameter, as depicted in Fig.~\ref{degree}.

 \section{Conclusion}
 \label{sec:conclusion}

 We have shown that electron spin dynamics in bichromatic KD scattering are very sensitive to the polarization states of the laser beams, whose frequency ratio was chosen as 2. This way, previous investigations of the subject \cite{Ahrens2012, Batelaan_Spin, Erhard2015, Dellweg2016, Freimund2003} have been extended. 
 
 When the fundamental laser beam is circular-polarized, no spin effects were found to leading order. The latter corresponds to a three-photon process, where two photons of the fundamental frequency are absorbed and a second-harmonic photon is emitted (or vice versa). However, at higher field intensities, where a perturbative description does not apply, spin-flip transitions can occur. They exhibit characteristic Rabi oscillations as a function of the interaction time. This effect was explicitly demonstrated for the configuration where both laser waves are corotating circular-polarized.
 
 Spin dynamics can be controlled very effectively when a second-harmonic beam of circular polarization is combined with a fundamental laser beam of linear polarization. This setup, which works at low field intensities, can be used as a longitudinal spin filter for electrons moving along the beam axis
 (similarly to the scheme proposed in \cite{Ahrens2016}). Transverse momentum components of the electrons would exert a detrimental impact, though, allowing for spin-independent scattering events.
 
 The setup can be optimized by utilizing a fundamental laser beam with a small degree of ellipticity. If the latter is properly chosen with $\approx 20$\%, the spin filter is effective for electrons with nonzero transverse momentum, as well. From an incident electron beam satisfying the Bragg condition, only, say, spin-up electrons are scattered (and thereby flipped to the spin-down state with respect to the laser beam axis) whereas spin-down electrons go through undeflected. The laser field configuration thus acts as a spin polarizer, splitting the electron beam according to its spin composition and creating two outgoing partial beams with parallel spin orientation. For comparison, we note that the interferometric spin polarizer proposed in \cite{Dellweg2016a} produces outgoing partial beams with opposite spin orientations along the magnetic field direction, in close analogy with the Stern-Gerlach effect \cite{Stern-Gerlach}.
 
 In summary, bichromatic KD scattering enables spin polarization of free electron beams. By properly chosing the polarization geometry of the laser waves, the spin dynamics can be controlled in order to optimize the desired spin properties of the output beams. \\

\section*{Acknowledgement}
This study was supported by SFB TR18 of the German Research Foundation (DFG) under project No. B11.


\begin{thebibliography}{100}
\bibitem{sigma-plus} If the weak mode had opposite sense of rotation, 
$\sigma_+$ would appear in the Rabi matrix, inducing oscillations 
between the states $|-2,\downarrow\rangle$ and $|2,\uparrow\rangle$.
\end{thebibliography}
\end{document}